\documentclass[pra,letterpaper,twocolumn,raggedbottom,nobalancelastpage,floatfix]{revtex4}
\usepackage{amsmath}
\usepackage{bm}
\usepackage{graphicx,subfigure,color}

\begin{document}

\title{Ultrahigh Purcell factors and Lamb shifts using slow-light
metamaterial waveguides}
\author{Peijun Yao, C. Van Vlack, A. Reza, M. Patterson, M. M. Dignam, and
S. Hughes}
\email{shughes@physics.queensu.ca}

\begin{abstract}
Employing a medium-dependent quantum optics formalism and a Green function solution of
Maxwell's equations, we study the enhanced spontaneous emission factors
(Purcell factors) and Lamb shifts from a quantum dot or atom near the
surface of a 
slow-light metamaterial waveguide. Purcell factors of approximately 250 and
100 are found at optical frequencies for $p-$polarized and $s-$polarized
dipoles respectively placed 28\thinspace nm (0.02\thinspace $\lambda _{0}$)
above the slab surface, including a realistic metamaterial loss factor of $\gamma /2\pi
=2\,\mathrm{THz}$. For smaller loss values, we demonstrate that the slow-light regime of odd
metamaterial waveguide propagation modes can be observed and
related to distinct resonances in the Purcell factors. Correspondingly, we predict
unusually large and rich Lamb shifts of approximately $-1$\thinspace GHz to $%
-6$\thinspace GHz for a dipole moment of 50 Debye.  We also make a direct calculation of the far field
emission spectrum 
which contains direct measurable access to these enhanced Purcell factors
and Lamb shifts.
\end{abstract}

\maketitle

%

{\catcode`\|=\active  \gdef\Braket#1{\left<\mathcode`\|"8000\let|\BraVert {#1%
}\right>}}

{\catcode`\|=\active  \gdef\set#1{\mathinner{\lbrace\,{\mathcode`\|"8000%
\let|\midvert #1}\,\rbrace}} \gdef\Set#1{\left\{\:{\mathcode`\|"8000\let|%
\SetVert #1}\:\right\}}}


\address{Department of Physics, Queen's University\\
Kingston, ON  K7L 3N6 Canada}




\section{Introduction}

Early in 1968, Veselago predicted that a planar slab of negative-index
material (NIM), which possesses both negative permittivity $\varepsilon $
and negative permeability $\mu $, could refocus electromagnetic waves~\cite%
{veselago}. While an interesting prediction, these so-called metamaterials
did not receive much attention in optics research until Sir John Pendry
showed that NIMs could be used as a \textquotedblleft
superlens\textquotedblright , which could overcome the diffraction limit of
conventional imaging system~\cite{pendry}. Subsequently, a host of
applications has been proposed, ranging from designs for optical cloaks to
hide objects~\cite{pendry2006},
through to schemes that completely stop light~\cite{kosmas}. Although most
of these schemes are idealized, and suffer in the presence of metamaterial
loss, they have nevertheless motivated significant experimental progress.
For example, Schurig \emph{et al}. experimentally demonstrated
some cloaking features by split-ring resonators~\cite{schurig}; and recent
achievements in fabrication have facilitated the realization of negative
indices at communication wavelength~\cite{dolling}, with some extension to
quasi-3D structures also reported~\cite{valentine}.

While it is certainly becoming established that metamaterials posses some remarkable
classical optical properties, less well studied are their quantum optical
properties, such as what happens to the spontaneous emission of an embedded
atom or quantum dot. In 1946, Purcell pointed out that due to the spatial
variation of the local photon density of states (LDOS), the spontaneous
emission rate in a cavity can be enhanced or suppressed depending upon the
distance between the mirrors~\cite{purcell}. The modification of spontaneous
emission due to inhomogeneous structures is a large field in its own right,
leading to applications in quantum optical technology~\cite{singlePhotons}.
In the domain of metamaterials, K\"{a}stel and coworkers investigated the
spontaneous emission of an atom placed in front of a mirror with a layer of
NIM~\cite{kastel}; this study was motivated by the perfect lens prediction
of a vanishing optical path length between the focal points, leading to the
peculiar property that the evanescent waves emerging from the source are
exactly reproduced; consequently they found that the spontaneous emission
can be completely suppressed. This prediction does not account for the
essential inclusion of metamaterial loss or absorption; yet, it is 
well known that metamaterials must be dispersive and absorptive to satisfy
the fundamental principle of causality and the Kramers-Kr\"{o}nig relation~%
\cite{stockman}. Not surprisingly, when absorption is necessarily taken into
account, then the predicted properties of an ideal lossless metamaterial can
qualitatively change. For example, the superlens and the invisible cloak are
never perfect~\cite{smith,yao}, and slow light modes can never be really
stopped and are usually impractically lossy~\cite{arvin}. Similarly, it is expected that absorption will have an important
influence on the modification of spontaneous emission~\cite{sambale}. In
this regard, Xu \emph{et al}. have extended the works of K\"{a}stel \emph{et
al}. to one dimensional right-handed and left-handed material layers and
find that nonradiative decay and instantaneous radiative decay will
certainly weaken the predicted inhibition of spontaneous emission~\cite%
{xu2006,xu2007}.

Some of the first theories to treat quantum electrodynamics near a interface were
introduced around 1984 by Wylie and Sipe~\cite{SipePRA1984,SipePRA1985},
where, using Green function techniques, they showed that the scattered field can be expressed in terms of the appropriate
Fresnel susceptibilities. Using such methods, it is now well known that
the photonic LDOS can be increased near a metallic surface, e.g. see Ref.~%
\cite{JoulainPRB2003}, whereby the $p-$polarized dipole couples to a
transverse magnetic (TM) surface plasmon polariton (SPP). Typically
such resonances are far from the optical frequency domain and they are
restricted to TM polarization; in addition, the emission is dominated by
quenching or non-radiative decay. Even in the presence of gratings, enhanced
emission via SPPs is not very practical~\cite{SunAPL2007}. However, the rich
waveguide properties of metamaterials have quite different polarization
dependences and mode structures than SSPs at a metal surface; for example,
they can support slow-light, bound propagation modes and transverse electric
(TE or $s-$polarized) SPPs. \ It is therefore of fundamental interest to
explore the quantum optical aspects of these novel waveguides.

Enhanced emission at the surface of both metals and metamaterial slabs was
studied in 2004 by Ruppin and Martin~\cite{RuppinMartin2004}. \ They noted
that resonance peaks due to $s-$polarized surface modes and waveguide modes
can appear for the metamaterial case, although they did not discuss the
origin of the waveguide peaks. Similar findings were found by Xu \emph{et al.}%
~\cite{XuPRA2009}, but the role of loss was not explored in detail but
rather it was treated as a perturbation and assumed to lead to only
dissipation; such an assumption 
is highly suspect in a NIM waveguide, since the entire modal characteristics
of the structure depend intimately upon the material loss and dispersion
profiles \cite{arvin}. Very recently, spontaneous emission enhancements in
NIM materials and interesting quantum interference effects have been reported by Li
\emph{et al.}~\cite{LiPEB2009}, although unrealistically small losses were typically
assumed, and again the physics behind the enhancement factors was not made
clear. In all of these works, there has been no quantitative connection made
to the complex band structure or to the far-field (and thus measurable)
spontaneous emission spectra or dipolar frequency shifts (Lamb shifts).


In this work, the modified spontaneous emission of a quantum dot or atom
(single photon emitter) situated above a slow-light metamaterial waveguide
is investigated in detail by employing a medium-dependent Green
function theory and comparing with the lossy guided waveguide modes. We
compute the PF as well as the spontaneous emission spectrum in the far field
by developing and using a rigorous quantum optics theory. We stress
that the recent prediction of completely stopped waveguide modes in a
metamaterial waveguide~\cite{kosmas} would lead to an infinite PF,
but as reported by Reza~\emph{et al}.~\cite{arvin}, the properties of the
slow-light modes are significantly changed in the presence of loss;
thus we also investigate the dependence on loss in some detail.
We show that the emission properties of a photon emitter can
act as a probe for below light-line propagation mode characteristics,
showing measurable enhanced radiative broadening and quantum Lamb shifts.
The Lamb shifts are found to be extremely rich and pronounced.
We also compare and contrast
these NIM quantum optical features with well known results for metallic
surfaces.

Our paper is organized as follows. In Sect.~\ref{modes}, we introduce the
NIM waveguide structure of interest and compute and discuss the band
structure for both $s$ (TE) and $p$ (TM) polarization. In Sect.~\ref{theory}%
, we present a rigorous theory for calculating the Purcell factor (PF), Lamb
shift, and emitted spectrum from a single photon emitter. From this theory,
we derive an explicit and analytical solution to the emitted field at any
spatial location using a full non-Markovian theory which is valid for any
general media (lossy and inhomogeneous). In Sect.~\ref{MultiGFT}, we discuss
a general technique for computing the multi-layered Green function using a
stratified medium technique of Paulus \emph{et al.}~\cite{Paulus-PRE-62-5797}%
, and formally separate the total Green function into a homogeneous and
scattered part to properly obtain the photonic Lamb shift. In Sect.~\ref%
{Calculation}, we present calculations for the Purcell effect and Lamb
shift, as well as the spontaneous emission spectra emitted into the far
field. Finally, we give our conclusions in Sect.~\ref{Conc}.

\section{Metamaterial slab waveguides: complex band structure and slow-light
propagation modes}

\label{modes}

\begin{figure}[tbp]
\includegraphics[width=0.4\textwidth]{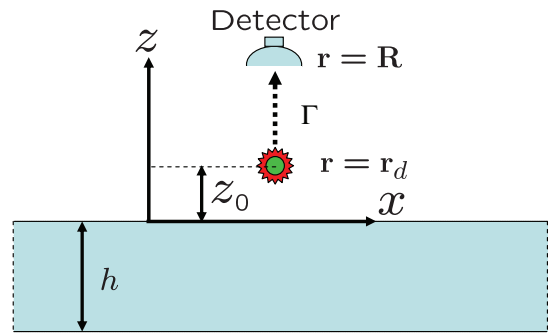}
\caption{Schematic diagram of the system being investigated. \ The green dot
at ${\bf r}={\bf r}_d$ refers to the atom or quantum dot that is decaying
radiatively (with a rate $\Gamma$), at a distance $z_{0}$ above the slab of thickness $h$.
The vertical position of the dot, $z_d=z_0$.}
\label{fig:schematic}
\end{figure}

In the following sections, we wish to calculate the spontaneous emission
from dipoles $\mathrm{\mathbf{d}}=d{\mathbf{e}}_{\alpha }$, where $\alpha =z$
or $\alpha =x(y)$. \ Because the spontaneous emission is strongly
affected by the metamaterial waveguide modes, in this section we present the
results of calculations of the dispersion of both TE and TM propagation
modes. \ The schematic diagram of the system under study is shown in Fig.~%
\ref{fig:schematic}.
The negative index slab is surrounded by air and assumed infinite (or much
larger than a wavelength) in the $x$ and $y$ directions.
The thickness of the slab is $h=280\,$nm. In view of the importance of
dispersion and absorption, we take both into account from the beginning,
which ensures causality and thus avoids unphysical results. The
dispersion is introduced via the Lorentz and Drude models~\cite{taflove}:
\begin{eqnarray}
\mu (\omega ) &=&1+\frac{\omega _{pm}^{2}}{\omega _{0}^{2}-\omega
^{2}-i\gamma \omega }, \\
\varepsilon (\omega ) &=&1-\frac{\omega _{pe}^{2}}{\omega ^{2}+i\gamma
\omega },
\end{eqnarray}%
where $\omega _{pm}$ and $\omega _{pe}$ are the magnetic and electric
plasmon frequencies, and $\omega _{0}$ is the \emph{atomic} resonance
frequency. In what follow, we are interested in waveguides with guided modes
at optical frequencies and thus take $\omega _{0}/2\pi =189.4\,\mathrm{THz}$%
, $\omega _{pm}/2\pi =165.4\,\mathrm{THz}$, $\omega _{pe}/2\pi =490\,\mathrm{%
THz}$.

To solve the complex band structure, one can work with a complex wave vector
($\beta $) and a real frequency ($\omega $), or, alternatively, a complex$%
-\omega $ and a real$-\beta $. The former is perhaps more appropriate for
modeling  plane wave excitation, while the latter is better
suited for a broadband excitation response that is invariant in $z$. Neither
of these approaches constitutes a complete connection to a broadband dipole
response, and thus we will show both solutions, and briefly discuss their
main features. The details of these two approaches for modelling
metamaterial waveguide properties will be presented elsewhere.

\begin{figure}[tbp]
\includegraphics[width=0.47\textwidth]{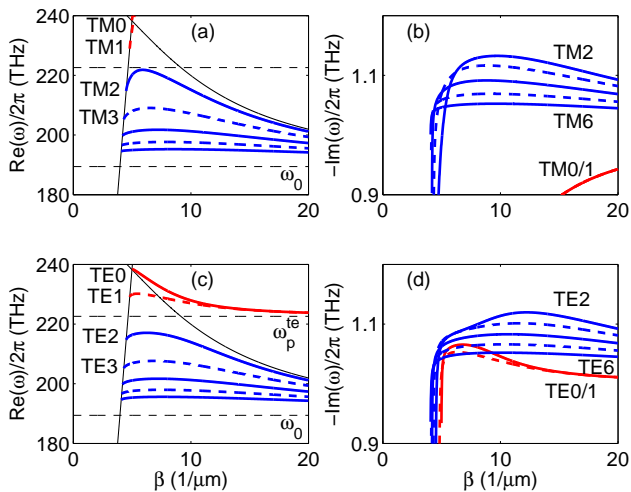}
\caption{ Dispersion curves of the lossy metamaterial waveguide for the
first few lower-order modes using the complex-$\protect\omega $ approach for
$\protect\gamma /2\protect\pi =2\,$THz ($\protect\gamma /\protect\omega %
_{0}\approx 0.01$). The red and blue curves show the surface plasmon
polariton modes and bound propagation modes, respectively. \ The solid and
dashed curves represent the even and odd modes, respecitvely. The solid
thick black curves display the vacuum light-line and metamaterial
light-line, while the horizontal dashed lines indicate the atomic ($\protect%
\omega _{0}$) and TE plasmon resonances ($\protect\omega _{p}^{\mathrm{te}}$%
). (a) $\mathrm{Re(\protect\omega )}$ versus $\protect\beta $ for TM modes.
The modes become more dense near the resonance frequency $\mathrm{\protect%
\omega _{0}}$, and form a continuum. (b) $\mathrm{Im(\protect\omega )}$
versus $\protect\beta $ for TM modes. (c)(d) The same as (a)(b) but for TE
modes. }
\label{fig:band_structure_complex_omega}
\end{figure}

\begin{figure}[tbp]
\includegraphics[width=0.44\textwidth]{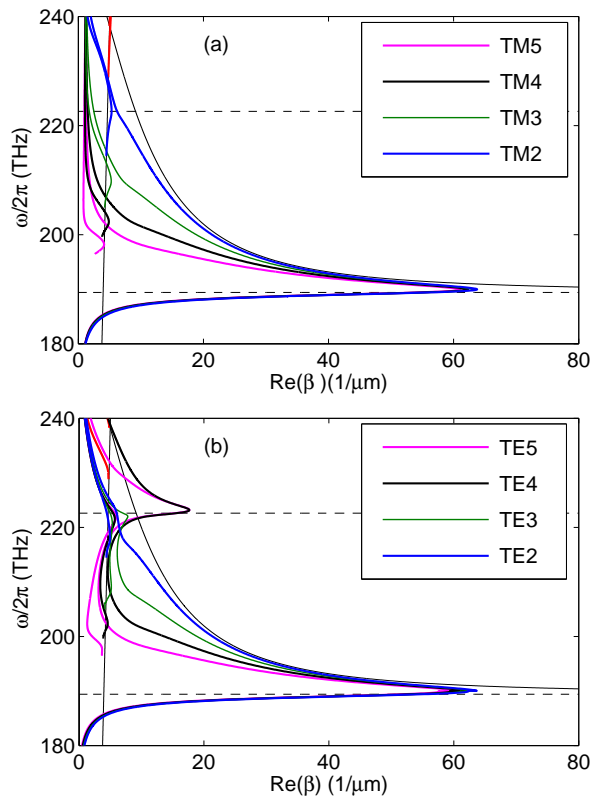}
\caption{Dispersion curves of the lossy metamaterial waveguide for the first
few lower-order modes in using the complex-$\protect\beta $ approach, where
only $\mathrm{Re}[\protect\beta ]$ is shown. (a) TM modes. (b) TE modes. The
thin solid black curves represent vacuum light-line and metamaterial
light-line. Note that formally labelling the curves is more difficult than in
Fig. 2 due to the merging of the different curves (e.g. the TM1 and TM5
curves near 220 THz).}
\label{fig:band_structure_complex_beta}
\end{figure}

The \emph{complex} dispersion curves of the aforementioned metamaterial
waveguide for both TE and TM modes are shown in Fig.~\ref%
{fig:band_structure_complex_omega} and Fig.~\ref%
{fig:band_structure_complex_beta}, using complex-$\omega $ and complex-$%
\beta $ approaches, respectively. In the complex$-\beta $ approach, we only
show the $\mathrm{Re}[\beta ]$ solution (Fig.~\ref%
{fig:band_structure_complex_beta}), as the $\mathrm{Im}[\beta ]$ simply
demonstrates the large losses in the regime of slow light~\cite{arvin}. These curves
come from the complex solution to the transcendental dispersion equation
derived from the Maxwell's equations and the guidance conditions~\cite{key-1}%
.

In Fig.~\ref{fig:band_structure_complex_omega}, we show the
first few TM (TE) propagation modes: TM2-TM6 (TE2-TM6), as well as TE0-TE1
(TM0-TE1) SPP modes (red curves); a wide range of frequencies is displayed
from from 180 to 240\thinspace THz. We stress that the \emph{TE-polarized}
SPPs are unique to metamaterials and the properties of these SSP modes can
be engineered to have a resonance 
in the optical frequency regime, 
near the propagation modes. For the TM case, there is also a
higher-lying SPP resonance, which we do not show as it is far outside the
frequency of interest ($\omega _{sp}^{\mathrm{tm}}\sim 346$\thinspace THz \, or specifically $\omega_{pe}/\sqrt{2}$); the TM0 and TM1 modes show the start  of the SPP
modes just below the air light line.

The properties of these metamaterial waveguide modes shown in Fig.~\ref{fig:band_structure_complex_omega} are considerably
different from those in a conventional dielectric waveguide~\cite{key-2}.
The most important difference is that
backward and forward propagating modes exist and, in a lossless
metamaterial waveguide, even stopped-light modes (where the slope goes to zero) are supported. However,
the necessary inclusion of intrinsic loss in a metamaterial slab
dramatically changes the dispersion curves near slow-light regions (see also
Ref.~\cite{arvin}). 
Thus one must include material losses to have any confidence in the results
and predictions. In the complex$-\omega $ solutions (Fig.~2), although the
slope and thus the group velocity $v_{g}$ is zero at some points, the use of the group velocity as a measure
of energy transport is not meaningful~\cite{arvin}, and there
is a finite imaginary part of frequency.
We note, however, that at the points where the slope is zero in Fig.
\thinspace \ref{fig:band_structure_complex_omega}(a) and (c), the
density of modes is large and as we shall see, this can lead to a
large Purcell factor at the relevant frequencies.  We note in particular
that because the slope of all of the different mode dispersions tends to
zero as $\beta$ goes to infinity, there is a large enhancement in the
density of states at $\omega = \omega_{0}$.

In Fig.~\ref{fig:band_structure_complex_beta} we show the dispersion curves
using the complex$-\beta$ approach.  In this approach, the impact of material losses
on the dispersion curves are much more drastic and these curves would appear to
have little to do with the curves in Fig.~\ref{fig:band_structure_complex_omega}.
However, despite their differences, there is a strong correspondence between the two
sets of curves. In a (fictitious) lossless waveguide, the two sets of curves would be
identical apart from some leaky modes that carry no energy.  In the complex$-\beta$ approach,
two degenerate leaky modes start at the zero-slope point of the non-SSP propagation modes
and move to higher frequencies.
In Fig.~\ref{fig:band_structure_complex_omega}, these leaky modes split and merge
seamlessly into the propagation mode dispersion, creating the split-curve structure
that is seen, e.g. for the TE4 mode near $\omega /2\pi = 201\,$THz. Although
the slope no longer goes to zero in the complex$-\beta$ approach at the point
where the propagation
mode splits into the two leaky modes, the energy velocity (which
is the correct measure of energy transport in a lossy system)
is quite small but nonzero.
Another important difference between the modes in the complex$-\beta$ approach and the complex-$\omega$
approach is that for the modes in the complex$-\omega$ approach,
$\beta \rightarrow \infty $ as $\omega \rightarrow \omega _{0}$, while in the
complex$-\beta$ approach, the modes bend back on themselves
at the atomic resonance frequency $\omega _{0}$.
Note that the group velocity, which is given by the slope of dispersion
curve, is infinite at this point. 
There is no paradox here because, as discussed above, the group velocity is
ill-defined in a lossy waveguide structure and the energy velocity is a
correct measure of the transport speed~\cite{key-4}. 
We have calculated the energy velocity and find that the minimum {\em energy}
velocity occurs exactly at the resonance frequency $\omega_{0}$, but that it is never zero;
for example, the energy velocity minimum for both TE3 and TM3 modes is found
to be around $10^{-3}c$ to $10^{-4}c$, where $c$ is the vacuum speed of
light. A similar effect occurs for TE modes near the plasmon resonance, where
in the in the complex$-\beta$ picture, the SSP modes not only penetrate below
the plasmon frequency, but they transform seamlessly into  the
higher-order leaky modes.  For example, the TE4 and TE5 modes merge into
SSP modes.  Interestingly, in Fig.~3(b), we also observe new resonances in the
left branch of the leaky TE
propagation modes, which couple to the TE SPP modes just below the bare SPP
resonance. \ Thus, two resonances  appear in this SPP frequency
regime; evidently, this is not expected from the complex$-\omega $
perspective.

Later, we will show that the resonant frequency and the slow
light regimes of the propagation modes 
are exactly coincident to that of the LDOS peaks in the spontaneous emission
spectrum, and thus both complex$-\omega $ and complex$-\beta $ band
structures are useful to gain insight into the origin of the LDOS peaks.
The spatial symmetries of the even and odd modes will also prove to be very
important; since the odd modes have a much larger
field amplitude at the surface, they couple much more strongly to a quantum
dot or atom near the
NIM surface.

\section{Quantum theory of spontaneous emission in metamaterial}

\label{theory}

\subsection{Purcell factor (PF) and Lamb shift}

The PF is a measure of the spontaneous emission rate enhancement; it is
defined as $PF={\Gamma }/{\Gamma _{0}}$, where $\Gamma $ (the Einstein
A coefficient) is the spontaneous emission rate associated with population
decay rate from an excited state to the ground state, and $\Gamma _{0}$ is
the spontaneous emission rate in vacuum or a lossless homogeneous medium. We
consider a system consisting of a quantum dot embedded in or near a general
dispersive, absorptive, and inhomogeneous medium. Employing a quantization
scheme that rigorously satisfies the Kramers-Kr\"{o}nig relations, and using
the electric dipole approximation,
an appropriate Hamiltonian -- following the works
of Welsch and coworkers --
can be written as~\cite{dung1,dung2,dungpra043816}:
\begin{eqnarray}
H &=&\!\!\hbar \omega _{d}\hat{\sigma}^{+}\!\hat{\sigma}^{-}+\sum_{\lambda
=e,m}\int d\mathbf{r}\int_{0}^{\infty }\!\!d\omega _{l}\hbar \omega _{l}\hat{%
\mathbf{f}}_{\lambda }^{\dagger }(\mathbf{r},\omega _{l})\cdot \hat{\mathbf{f%
}}_{\lambda }(\mathbf{r},\omega _{l})  \notag \\
&-&[\hat{\sigma}^{+}\mathbf{d}+\hat{\sigma}^{-}\mathbf{d}]\cdot \mathbf{F}(%
\mathbf{r}_{d}),
\end{eqnarray}%
where $\hat{\mathbf{f}}_{\lambda }^{\dagger }(\mathbf{r},\omega _{l})$ and $%
\hat{\mathbf{f}}_{\lambda }(\mathbf{r},\omega _{l})$ are the continuum
bosonic-field operators of the electric ($\lambda =e$) and magnetic field ($%
\lambda =m$) with eigenfrequency $\omega _{l}$, $\hat{\sigma}^{(\pm )}$ are
the Pauli operators of the electron-hole pair (exciton), and $\omega _{d}$
and $\bm{d}=\bm{n}_{d}d$ ($d$ is assumed to be real) are the transition
frequency and dipole moment of the dot, respectively. \ The field operator $%
\hat{\mathbf{F}}$
is essentially the electric field operator augmented by the quantum dot
polarization and can be expressed as $\hat{\mathbf{F}}=\hat{\mathbf{D}}/
({\varepsilon _{0}\varepsilon })+{\hat{\mathbf{P}}}/({\varepsilon
_{0}\varepsilon })$, where ${\hat{\mathbf{D}}}$
is the displacement field (${\hat{\mathbf{E}}} =\hat{\mathbf{D}}/({\varepsilon _{0}\varepsilon })$ is the electric field away from the dipole),
${\varepsilon }$ is the complex relative permittivity and ${\hat{\mathbf{P}}}$
is the polarization arising from the quantum dot dipole;
this latter contribution is needed as it is the displacement field that
should couple to the dipole in the interaction Hamiltonian~\cite%
{HealyPRA1980,PowerPRA1982,SipePRA1984}.
Using the above formalism, we derive
\begin{eqnarray}
\hat{\mathbf{F}}(\mathbf{r},t)& \!\!= &i\sqrt{\frac{\hbar}{\pi\varepsilon_0}}\!\!\int_0^{\infty}\!\!\!d\omega_l\!\!\!\int
\!\!d\mathbf{r}'\mathbf{G}(\mathbf r,\mathbf
r';\omega_l)\cdot\left [\sqrt{\varepsilon_I(\mathbf{r}',\omega_l)}
 \right . \nonumber \\
&\hbox{} & \left . \!\!\!\!\!\!\!\!\hat{\mathbf{f}}_e(\mathbf{r}',\omega_l;t)+
\frac{c}{\omega_l}\nabla\times\sqrt{-\kappa_I(\mathbf{r}',\omega_l)}\,\hat{\mathbf{f}}_m(\mathbf{r}',\omega_l;t)\right]
\nonumber \\
&\hbox{} & \!\!\!\!\! + H.c. + \frac{{\bf d}[\hat{\sigma}^+(t) + \hat{\sigma}^-(t)]\,\delta({\bf r}-{\bf r}_d)}
{\varepsilon_0\varepsilon({\bf r})},
\end{eqnarray}
where the last term represents the polarization field from the dipole, $%
\varepsilon _{I}(\mathbf{r},\omega _{l})$ and $\kappa _{I}(\mathbf{r},\omega
_{l})$ are the imaginary parts of $\varepsilon (\mathbf{r},\omega _{l})$ and
$1/\mu (\mathbf{r},\omega _{l})$ respectively, and $\varepsilon (\mathbf{r}%
,\omega _{l})$ and $\mu (\mathbf{r},\omega _{l})$ are the relative complex
permittivity and permeability. The dyadic $\mathbf G(\mathbf r, \mathbf r';\omega_l)$
is the
electric-field Green function (GF) that 
describes the field response at
$\mathbf r$ to an oscillating
polarization dipole at $\mathbf r'$ as a function of frequency;
the GF is defined from
\begin{eqnarray}
\!\!\!\!\!\!\!\! \left[{ \nabla}\!\times \! \frac{1}{\mu(\omega_l,\mathbf r)}
{\nabla}\!\times\! -\frac{\omega_l^2}{c^2}\varepsilon(\omega_l,\mathbf r)\right]
\textbf{G}(\mathbf r,\mathbf r';\omega_l) = \nonumber \\
  \frac{\omega_l^2}{c^2}{\bf 1}\,\delta({\bf r}-{\bf r}') ,
\end{eqnarray}
where ${\bf 1}$ is the unit tensor.

From the above Hamiltonian, we derive the Heisenberg equations of motion for
the time-dependent operator equations as ($t$ is implicit):
\begin{eqnarray}
&&\frac{d\hat{\sigma}^-}{dt}=-i\omega_l \hat{\sigma}^-+i\hbar^{-1} \mathbf{d}
\cdot \hat{\mathbf{F}} (\mathbf{r}_d), \\
&&\frac{d\hat{\mathbf{f}}_e(\mathbf{r},\omega_l)}{dt}=-i\omega_l \hat{%
\mathbf{f}}_e(\mathbf{r},\omega_l) \notag \\
&&\qquad\qquad-\sqrt{\frac{\varepsilon_I(\mathbf{r},\omega_l)}{%
\pi\hbar\varepsilon_0}} \bm{d}\cdot\mathbf{G}^*(\mathbf{r}_d,\mathbf{r}%
;\omega_l)[\hat{\sigma}^- + \hat{\sigma}^+], \\
&&\frac{d\hat{\mathbf{f}}_m(\mathbf{r},\omega_l)}{dt}=-i\omega_l \hat{%
\mathbf{f}}_m(\mathbf{r},\omega_l)  \notag \\
&& \!\!\!\!\!\!\!\!\!\!\!\!\!\!\!\! \qquad\qquad-\sqrt{\frac{-\kappa_I(%
\mathbf{r},\omega_l)}{\pi\hbar\varepsilon_0}} \frac{c}{\omega_l}\bm{d}\cdot[%
\mathbf{G}^*(\mathbf{r}_d,\mathbf{r};\omega_l)\times\nabla_{}][\hat{\sigma}%
^- + \hat{\sigma}^+] ,  \notag \\
\end{eqnarray}
where $[\mathbf{G}^*(\mathbf{r}_d;\mathbf{r};\omega_l)\times\nabla_{}]_{ij}=%
\epsilon_{jkl}\partial_k {G}_{il}^*(\mathbf{r}_d,\mathbf{r};\omega_l)$, and
we have used the one photon or weak excitation approximation through $%
\hat\sigma_z\hat{\mathbf{F}}=-\hat{\mathbf{F}}$. We can make a Laplace
transform on the above set, defined through $\hat O_i(\omega) = \int_0^\infty
e^{i\omega t} \hat O_i(t) dt$, and obtain
\begin{eqnarray}
\hat{\sigma}^-(\omega)&=&\frac{i\hat{\sigma}^-(t=0)} {\omega-\omega_d}-\frac{%
\hbar^{-1}\mathbf{d}\cdot\hat{\mathbf{F}} (\mathbf{r}_d,\omega)}{%
\omega-\omega_d},  \label{eq:f1} \\
\hat{\sigma}^+(\omega)&=&\frac{i\hat{\sigma}^+(t=0)} {\omega+\omega_d}+\frac{%
\hbar^{-1}\mathbf{d}\cdot\hat{\mathbf{F}} (\mathbf{r}_d,\omega)}{%
\omega+\omega_d},  \label{eq:f2} \\
\hat{\mathbf{f}}_e(\mathbf{r},\omega_l;\omega)&=& \hat{\mathbf{f}}_e^0(%
\mathbf{r},\omega_l;\omega)  \notag \\
& \hbox{} & \!\!\!\!\!\!\!\!\!\!\!\!\!\!\! -\sqrt{\frac{\varepsilon_I(%
\mathbf{r},\omega_l)}{\pi\hbar\varepsilon_0}} \mathbf{d}\cdot\mathbf{G}^*(%
\mathbf{r}_d,\mathbf{r};\omega_l)\frac{i[\hat{\sigma}^-(\omega)+\hat{\sigma}%
^+(\omega)]} {\omega-\omega_l} ,  \notag  \label{eq:f3} \\
\\
\hat{\mathbf{f}}_m(\mathbf{r},\omega_l;\omega)&=& \hat{\mathbf{f}}_m^0(%
\mathbf{r},\omega_l;\omega)  \notag \\
&\hbox{}& \!\!\!\!\!\!\!\!\!\!\!\!-\sqrt{\frac{-\kappa_I(\mathbf{r},\omega_l)%
}{\pi\hbar\varepsilon_0}} \frac{c}{\omega_l} \mathbf{d}\cdot[\mathbf{G}^*(%
\mathbf{r}_d,\mathbf{r};\omega_l)\times\nabla_{}]  \notag \\
&\hbox{}& \frac{i[\hat{\sigma}^-(\omega)+\hat{\sigma}^+(\omega)]}{%
\omega-\omega_l} ,  \label{eq:f4}
\end{eqnarray}
where $\mathbf{f}^0$ represents a possible \emph{free field} or homogeneous
driving field in the absence of any quantum dot or atom.

We next assume that the initial field is the vacuum field (${i.e.}$, $%
\hat{\mathbf{f}}_{\lambda }^{0}(\mathbf{r},\omega _{l};\omega )=0$),
substitute Eqs.(\ref{eq:f3}-\ref{eq:f4}) into Eq.(4), and make use of the
relation (see \cite{dungpra043816}):
\begin{eqnarray}
\int d\mathbf{s}\{-\kappa_I[\mathbf G(\mathbf{r},\mathbf{s};\omega_l)\times \nabla_s][\nabla_s\times {\bf G}^*(\mathbf{s},\mathbf{r'};\omega_l)]\frac{c^2}{\omega_l^2}+\nonumber\\
\varepsilon_I(\mathbf{s},\omega_l)\mathbf G(\mathbf{r},\mathbf{s},\omega_l)
\cdot {\bf G}^*(\mathbf{s},\mathbf{r'};\omega_l)\}=\rm
Im \mathbf G(\mathbf{r},\mathbf{r'};\omega_l).\ \
\end{eqnarray}
Subsequently, we obtain an explicit solution for the dipole operators (and thus the polarization operator):
\begin{eqnarray}
&&\hat{\sigma}^{-}(\omega )+\hat{\sigma}^{+}(\omega )=  \notag \\
&&\!\!\!\!\!-\frac{i\hat{\sigma}^{-}(t=0)(\omega +\omega _{d})+i\hat{\sigma}%
^{+}(t=0)(\omega -\omega _{d})}{\omega _{d}^{2}-\omega ^{2}-2\omega _{d}\,%
\mathbf{d}\cdot {[\mathbf{G}}(\mathbf{r}_{d},\mathbf{r}_{d};\omega )+\frac{%
\delta (\mathbf{r}-\mathbf{r}_{d})}{\varepsilon({\bf r})}]\cdot \mathbf{d}/\hbar
\varepsilon _{0}},  \notag \\
&&
\end{eqnarray}%
which we can rewrite as
\begin{eqnarray}
&&\hat{\sigma}^{-}(\omega )+\hat{\sigma}^{+}(\omega )=  \notag \\
&&\ \ \ \ \ -\frac{i\hat{\sigma}^{-}(t=0)(\omega +\omega _{d})+i\hat{\sigma}%
^{+}(t=0)(\omega -\omega _{d})}{\omega _{d}^{2}-\omega ^{2}-2\omega _{d}\,%
\mathbf{d}\cdot {\mathbf{K}}(\mathbf{r}_{d},\mathbf{r}_{d};\omega )\cdot
\mathbf{d}/\hbar \varepsilon _{0}},\ \ \ \ \
\end{eqnarray}%
where the new GF, $\mathbf{K}(\mathbf{r},\mathbf{r}^{\prime };\omega )\equiv
\mathbf{G}(\mathbf{r},\mathbf{r}^{\prime };\omega )+\delta (\mathbf{r}-%
\mathbf{r}^{\prime })/\varepsilon (\mathbf{r})$. This is exactly the same
form as the GF used in the formalism by Wubs \emph{et al.}~\cite{wubs04},
where the $\mathbf{K}$ function appears naturally when working with mode
expansion techniques for lossless inhomogeneous media. The origin of the
discrepancy between theories that use $\mathbf{G}$ or $\mathbf{K}$, is
because 
the correct interaction Hamiltonian should really contain a displacement
field~\cite{HealyPRA1980,PowerPRA1982,SipePRA1984},
as has been accounted for in our Eq.~(4). This subtlety becomes important,
e.g., when deriving a Lippman-Schwinger equation for the electric-field
operator, which can only be achieved through use of $\mathbf{K}$~\cite%
{wubs04}.

It is worth noting that the above equations are obtained with no Markov
approximation, so they can be applied to both weak and strong coupling
regimes of cavity-QED. In addition, we have made no rotating-wave
approximations. In the weak to intermediate coupling regime, the decay rate
of spontaneous emission $\Gamma$ can be conveniently expressed via the
photon GF through,
\begin{eqnarray}
\label{PF}
\Gamma({\mathbf r}_d, \omega)=\frac{2{\bf d}\cdot {\rm Im}[{\mathbf
K}({\mathbf r}_d, {\mathbf r}_d;\omega)] \cdot{\bf d}}{\hbar
\varepsilon_0} \ ,
\end{eqnarray}
where for free space, $\mathrm{Im}[\mathbf{K}^{\mathrm{vac}}(\omega)]=%
\mathrm{Im}[\mathbf{G}^{\mathrm{vac}}(\omega)]=\omega^3/6\pi c^3$, and so $%
\Gamma^0=2d^2\omega^3/(\hbar\varepsilon_06\pi c^3)$.

Within the dipole
approximation, the above formalism is exact, and 
for \emph{lossless} media, Eq.~(\ref{PF}) can be reliably applied as soon as
$\mathbf{G}$ is known, and one can exploit $\mathrm{Im}[\mathbf{G}(\mathbf{r}%
,\mathbf{r}^{\prime};\omega)] =\mathrm{Im}[\mathbf{K}(\mathbf{r},\mathbf{r}%
^{\prime};\omega)]$, since $\varepsilon$ is real. In a previous paper
dealing with lossless photonic crystals~\cite{shughes_optexpress}, two of us
adopted precisely $\mathbf{K}(\mathbf{r}%
,\mathbf{r}^{\prime};\omega)$, since it can be constructed in
terms of the transverse modes. However, for lossy structures such as metals
and metamaterials, 
$\mathrm{Im}[\mathbf{G}(\mathbf{r},\mathbf{r};\omega)]$ diverges~\cite{dung3}%
, so we are forced to confront the immediate unphysical consequences of a
dipole approximation. For both real and complex $\varepsilon$, $\mathrm{Re}[%
\mathbf{G}(\mathbf{r},\mathbf{r};\omega)]$ also diverges, which is well
known. These GF divergences, as $\mathbf{r}\rightarrow \mathbf{r}^{\prime}$,
are of course not physical and are simply a consequence of using the dipole
approximation. Any finite size emitter, no matter how small, will have a
finite PF and a finite Lamb shift~\cite{lamb}. The usual procedure for a
lossy homogenous structure is to either regularize the GF by introducing a
high momentum cut off~\cite{lagRev}, or to introduce a real or virtual
cavity around a finite size emitter and analytically integrate the
homogenous GF~\cite{dung4}.

In the remainder of this paper, we will only concern ourselves with dipole
emitters located in free space above a NIM waveguide; we will, however,
revisit the problem of dipoles inside a NIM in future work, where one must carefully
account for local field effects. For any \emph{%
inhomogeneous} structure such as a layered waveguide, a convenient approach
to using the GF is to formally separate it 
into two parts, namely a homogeneous contribution $\mathbf{G}^{\mathrm{hom}}$
(whose solution can be obtained analytically) and a scattering contribution $%
\mathbf{G}^{\mathrm{scatt}}$. This approach, which is especially well-suited
to dipoles in free space, is the approach that we will adopt below.
Using this separation, one can identify the non-divergent PF and Lamb shift
solely from the scattered part of the GF. We obtain
\begin{eqnarray}
\label{PF2}
\Gamma^{\rm scatt}({\mathbf r}_d, \omega)=\frac{2{\bf d}\cdot {\rm Im}[{\mathbf
G}^{\rm scatt}({\mathbf r}_d, {\mathbf r}_d;\omega)] \cdot{\bf d}}{\hbar
\varepsilon_0} \ ,
\end{eqnarray}
so that the Purcell factor, for a dot in free space, is
\begin{eqnarray}
\label{PF3}
PF({\bf r}_d) = 1 + \frac{\Gamma^{\rm scatt}({\mathbf r}_d, \omega)}{\Gamma^0(\omega)}.
\end{eqnarray}
Similarly, the Lamb shift is given by
\begin{eqnarray}
\label{Lamb}
\Delta\omega({\mathbf r}_d, \omega)=-\frac{{\bf d}\cdot {\rm Re}[{\mathbf
G}^{\rm scatt}({\mathbf r}_d, {\mathbf r}_d;\omega)] \cdot{\bf d}}{\hbar
\varepsilon_0} \ ,
\end{eqnarray}
where we have neglected the vacuum Lamb shift from the homogenous GF, since
it can be thought to exist altready in the definition of $\omega _{d}$, and
in any case it will be much smaller than any resonant frequency shifts coming
from $\mathbf{G}^{\mathrm{scatt}}$. While, in principle, one can apply mass
renormalization techniques to also obtain the vacuum (or electronic) Lamb
shift~\cite{milonni}, any observable shift will be related to -- and
completely dominated by -- the photonic Lamb shift, and thus from $\mathbf{G}%
^{\mathrm{scatt}}$. 

\subsection{Spontaneous emission spectrum}

Next, we will obtain an exact expression for the emitted spectrum. From
Eq.~(4) and Eqs.~(11-15), we obtain the analytical expression for the electric field
operator, $\hat{\mathbf{E}}\left( \mathbf{R},\omega \right) $, for $\mathbf{R}%
\neq \mathbf{r}_{d}$:
\begin{eqnarray}
\hat{\mathbf{E}}(\mathbf  R, \omega)&=&\int d\omega_l\frac{1}{\pi\varepsilon_0}{\rm Im}\mathbf{G}(\mathbf R, \mathbf r_d;\omega_{l})\cdot {\bf d} \, \frac{\hat{\sigma}^-(\omega)+\hat{\sigma}^+(\omega)}{\omega-\omega_l}\nonumber\\
&=&\frac{1}{\varepsilon_0}\mathbf{G}(\mathbf R, \mathbf r_d,\omega)\cdot {\bf d} [\hat{\sigma}^-(\omega)+\hat{\sigma}^+(\omega)],
\end{eqnarray}
where $\frac{i}{(\omega_l -\omega+i\epsilon _{+})}=\pi \delta (\omega_l
-\omega)+iP(\frac{1}{\omega_l -\omega})$ ($\omega_l$ is the integration variable) has been used. Note that the
principal value of the integral cannot be neglected, otherwise only the
imaginary part of GF in Eq.~(20) is retained. This can be contrasted to the
expression derived by Ochia \emph{et al.}~\cite{sakoda}, where the real part
of the GF was omitted because they neglected the principal value of
integral; however, this is unphysical and yields incorrect spectral shapes
in general.

The power spectrum of spontaneous emission can be obtained from $S(\mathbf{R}%
,\omega )=\int_{0}^{\infty }\int_{0}^{\infty }dt_{1}dt_{2}e^{i\omega
(t_{2}-t_{1})}\langle \mathbf{\hat{E}}^{-}(t_{1})\mathbf{\hat{E}}%
^{+}(t_{2})\rangle $, leading to 
$S(\mathbf{r},\omega )=\langle (\mathbf{\hat{E}}(\omega ))^{\dagger }\mathbf{%
\hat{E}}(\omega )\rangle $. Using Eq.~(20) and Eq.~(15), one has, again for $%
\mathbf{R}\neq \mathbf{r}_{d}$, 
\begin{equation}
S(\mathbf{R},\omega )=\left\vert \frac{\mathbf{d}\cdot \mathbf{G}(\mathbf{R},%
\mathbf{r}_{d};\omega )(\omega +\omega _{d})/\varepsilon _{0}}{\omega
_{d}^{2}-\omega ^{2}-2\omega _{d}\mathbf{d}\cdot {\bf G}^{\mathrm{\rm scatt}}({\bf r}_{d},%
\mathbf{r}_{d};\omega )\cdot \mathbf{d}/\hbar \varepsilon _{0}}\right\vert
^{2}.\ \ \
\end{equation}%
This is in an identical form to the emission spectrum derived for a
lossless material~\cite{shughes_optexpress}, showing that the electric-field
spectrum at
 $\mathbf r$
 depends on the two-space point GF, $\mathbf{G}(\mathbf{R},\mathbf{r}_d;\omega)$,
 which describes radiative propagation from  the dot position to the detector.
 All that remains to be done is obtain the GF, which we discuss next.

\section{Multi-layered Green function: Plane wave expansion technique}

\label{MultiGFT} The classical GF, $\mathbf{G}\left( \mathbf{r},\mathbf{r}%
^{\prime };\omega \right) $, describes the response of a system at the
position $\mathbf{r}$ to a polarization dipole located at $\mathbf{r}%
^{\prime }$, so that the total electric field 
$\mathbf{E}\left( \mathbf{r},\omega \right) =\mathbf{G}\left( \mathbf{r},%
\mathbf{r}^{\prime };\omega \right) \cdot \mathbf{p}\left( \mathbf{r}%
^{\prime },\omega \right) .$ 
In the case of a multilayer planar system~\cite%
{Paulus-PRE-62-5797,Tomas-PRA-51-2545}, when calculating the electric field
in the same layer as the dipole, and as mentioned earlier, it is possible to
formally write the GF in terms of a homogenous part and a scattered part.
Formally one has
\begin{equation}
\mathbf{G}\left( \mathbf{r},\mathbf{r}^{\prime };\omega \right) =\mathbf{G}^{%
\mathrm{hom}}\left( \mathbf{r},\mathbf{r}^{\prime };\omega \right) +\mathbf{G%
}^{\mathrm{scatt}}\left( \mathbf{r},\mathbf{r}^{\prime };\omega \right) \,.
\end{equation}%
Because we consider $z$ and $x(y)$ oriented dipoles separately, we only
require the diagonal elements of the GF above the slab and the GF tensor
elements can be greatly simplified. We take the source and field points to
be $\mathbf{r}=\left( \mathbf{\rho },z\right) $ and $\mathbf{r}^{\prime
}=\left( \mathbf{\rho },z^{\prime }\right) $, i.e. the transverse position, $%
\mathbf{\rho =}\left( x,y\right) $ of the dipole and observation points are
equal. We will use the following notation to label the three-layered
structure: region 1 is air, region 2 is metamaterial and region 3 is air.
For the total GF, when both $z$ and $z^{\prime }$ are in region 1, we have,
for $z\leq z^{\prime }$,
\begin{equation}
\begin{split}
G_{xx/yy}\left( \bm{\rho},z,z^{\prime },\omega \right) & =\frac{i\mu \left(
\mathbf{r}^{\prime },\omega \right) \omega ^{2}}{8\pi c^{2}}\int_{0}^{\infty
}dk_{\rho }k_{\rho } \\
& \!\!\!\!\!\!\!\!\!\!\!\!\left[ \frac{1}{k_{1z}}\left(
r_{1,-}^{s}e^{ik_{1z}\left( z+z^{\prime }\right) }+e^{-ik_{1z}\left(
z-z^{\prime }\right) }\right) \right. \\
& \!\!\!\!\!\!\!\!\!\!\!\!\left. -\frac{k_{1z}}{k_{1}^{2}}\left(
r_{1,-}^{p}e^{ik_{1z}\left( z+z^{\prime }\right) }+e^{-ik_{1z}\left(
z-z^{\prime }\right) }\right) \right] ,
\end{split}
\label{eq:gtot_xx_zl}
\end{equation}%
\begin{equation}
\begin{split}
G_{zz}\left( \bm{\rho},z,z^{\prime },\omega \right) & =-\frac{1}{\varepsilon
\left( \mathbf{r}^{\prime },\omega \right) }\delta \left( z-z^{\prime
}\right) +\frac{i\mu \left( \mathbf{r}^{\prime },\omega \right) \omega ^{2}}{%
4\pi c^{2}} \\
& \!\!\!\!\!\!\!\!\!\!\!\!\!\!\!\!\!\!\!\!\!\!\!\int_{0}^{\infty }dk_{\rho }%
\frac{k_{\rho }^{3}}{k_{1z}k_{1}^{2}}\left( r_{1,-}^{p}e^{ik_{1z}\left(
z+z^{\prime }\right) }+e^{-ik_{1z}\left( z-z^{\prime }\right) }\right) ,
\end{split}
\label{eq:gtot_zz_zl}
\end{equation}%
and with $z>z^{\prime }$,
\begin{equation}
\begin{split}
G_{xx/yy}\left( \bm{\rho},z,z^{\prime },\omega \right) =& =\frac{i\mu \left(
\mathbf{r}^{\prime },\omega \right) \omega ^{2}}{8\pi c^{2}}\int_{0}^{\infty
}dk_{\rho }k_{\rho } \\
& \!\!\!\!\!\!\!\!\!\!\!\!\!\!\!\!\!\!\!\left[ \frac{1}{k_{1z}}\left(
r_{1,-}^{s}e^{ik_{1z}\left( z+z^{\prime }\right) }+e^{ik_{1z}\left(
z-z^{\prime }\right) }\right) \right. \\
& \!\!\!\!\!\!\!\!\!\!\!\!\!\!\!\!\!\!\!-\left. \frac{k_{1z}}{k_{1}^{2}}%
\left( r_{1,-}^{p}e^{ik_{1z}\left( z+z^{\prime }\right) }-e^{ik_{1z}\left(
z-z^{\prime }\right) }\right) \right] ,
\end{split}
\label{eq:gtot_xx_zg}
\end{equation}%
\begin{equation}
\begin{split}
G_{zz}\left( \bm{\rho},z,z^{\prime },\omega \right) & =\frac{i\mu \left(
\mathbf{r}^{\prime },\omega \right) \omega ^{2}}{4\pi c^{2}}\int_{0}^{\infty
}dk_{\rho }\frac{k_{\rho }^{3}}{k_{1z}k_{1}^{2}} \\
& \left( r_{1,-}^{p}e^{ik_{1z}\left( z+z^{\prime }\right) }+e^{ik_{1z}\left(
z-z^{\prime }\right) }\right) \,.
\end{split}
\label{eq:gtot_zz_zg}
\end{equation}%
Here, for $s$ (TE) and $p$ (TM) polarization,
\begin{equation}
r_{1,-}^{\left( s/p\right) }=r_{12}^{\left( s/p\right) }+\frac{%
t_{12}^{\left( s/p\right) }t_{21}^{\left( s/p\right) }r_{23}^{\left(
s/p\right) }e^{2ik_{2z}h}}{1-r_{21}^{\left( s/p\right) }r_{23}^{\left(
s/p\right) }e^{2ik_{2z}h}}\,,
\end{equation}%
and the wave vector, $k_{l}=n\omega /c$ in medium $l$. Calculating the z
component of the wave vector as $k_{lz}=\pm \left( k_{l}^{2}-k_{\rho
}^{2}\right) ^{1/2}$ when $\mathrm{Re}(k_{l})>\mathrm{Re}(k_{\rho })$ where
the positive sign is for positive index materials and the negative sign is
for negative index materials. For $\mathrm{Re}(k_{l})<\mathrm{Re}(k_{\rho })$
we have $k_{lz}=i\left( k_{\rho }^{2}-k_{l}^{2}\right) ^{1/2}$ for both
positive and negative index materials~\cite{dungpra043816,Ramakrishna2005}.
The reflection and transmission coefficients are
\begin{eqnarray}
r_{ij}^{s} &=&\frac{\mu _{j}k_{iz}-\mu _{i}k_{jz}}{\mu _{j}k_{iz}+\mu
_{i}k_{jz}},\quad \quad r_{ij}^{p}=\frac{\varepsilon _{j}k_{iz}-\varepsilon
_{i}k_{jz}}{\varepsilon _{j}k_{iz}+\varepsilon _{i}k_{jz}} \\
t_{ij}^{s} &=&\frac{2\mu _{j}k_{iz}}{\mu _{j}k_{iz}+\mu _{i}k_{jz}},\quad
\quad t_{ij}^{p}=\frac{2\varepsilon _{j}k_{iz}}{\varepsilon
_{j}k_{iz}+\varepsilon _{i}k_{jz}}.
\end{eqnarray}

From the above expressions it is seen that the scattered GF for any $z$ in
region 1 can be written as
\begin{equation}
\begin{split}
G_{xx}^{\mathrm{scatt}}\left( \bm{\rho},z,z^{\prime },\omega \right) & =%
\frac{i\mu \left( \mathbf{r}^{\prime },\omega \right) \omega ^{2}}{4\pi c^{2}%
}\int_{0}^{\infty }dk_{\rho }k_{\rho } \\
& \!\!\!\!\!\!\!\!\!\!\!\!\!\!\!\!\!\!\left[ \left( \frac{r_{1,-}^{s}}{%
2k_{1z}}e^{ik_{1z}\left( z+z^{\prime }\right) }-\frac{k_{1z}r_{1,-}^{p}}{%
2k_{1}^{2}}e^{ik_{1z}\left( z+z^{\prime }\right) }\right) \right] ,
\end{split}
\label{eq:gscatt_xx}
\end{equation}%
\begin{equation}
\begin{split}
G_{zz}^{\mathrm{scatt}}\left( \bm{\rho},z,z^{\prime },\omega \right) & =%
\frac{i\mu \left( \mathbf{r}^{\prime },\omega \right) \omega ^{2}}{4\pi c^{2}%
}\int_{0}^{\infty }dk_{\rho } \\
& \frac{k_{\rho }^{3}}{k_{1z}k_{1}^{2}}r_{1,-}^{p}e^{ik_{1z}\left(
z+z^{\prime }\right) }\,.
\end{split}
\label{eq:gscatt_zz}
\end{equation}%
We highlight again that 
the difference between Eqs.~(\ref{eq:gtot_xx_zl}-\ref{eq:gtot_zz_zg}) and
Eqs.~(\ref{eq:gscatt_xx}-\ref{eq:gscatt_zz}) is simply the homogeneous GF~%
\cite{Tomas-PRA-51-2545}.

From a numerical perspective, the task is to
solve Sommerfeld integrals. Such equations can be integrated in the lower
half of the complex plane using the method described by Paulus \emph{et al.}~%
\cite{Paulus-PRE-62-5797} for positive index materials (where the poles are
in the first and the third quadrant of the complex plane), or in the upper
half of the complex plane for negative index materials (where the poles are
in the second and the fourth quadrant of the complex plane); this method
avoids any poles which may be near the real $k_{\rho }$ axis and improves
numerical convergence, though it is unnecessary for large material loss. For
our particular calculations, we integrated Eqs.~(\ref{eq:gtot_xx_zl}-\ref%
{eq:gscatt_zz}) using an adaptive Gauss-Kronrod quadrature which was
verified to be well converged for a relative tolerance of $10^{-4}$.
Specifically, the above equations were integrated along an elliptical path
around the region containing the bound and radiation mode contributions~\cite%
{Paulus-PRE-62-5797}, with the semi-major axis was $3\,\left\vert \mathrm{Re}%
\left( k_{2}\right) \right\vert /2$ and the semi-minor axis was $\left\vert
\mathrm{Re}\left( k_{2}\right) \right\vert /1000$. After integrating along
the elliptical path, the equations were integrated into the evanescent
region along the real $k_{\rho }$ axis. An additional advantage of this
technique is that the integrand contributions from the bound and evanescent
modes can be conveniently compared with the band structure, by examining the
individual $s-$ and $p-$ polarized contributions as a function of $k_{\rho }$
for a given $\omega $, where it becomes obvious that the full GF solution
requires both complex$-\omega $ and complex$-\beta $ pictures. Since the GF
approach may be termed the complete answer, it is clear that the band
structure approaches, either complex$-\omega $ or complex$-\beta $, merely
yield a limited sub-set solution about dipole coupling in these structures;
having said that, both approaches (band structure and GF) offer a clear
connection to the underlying physics of Purcell factors and Lamb shifts.

\section{Spontaneous Emission Calculations for a slow-light metamaterial
waveguide}

\label{Calculation}

\subsection{Enhancement of the spontaneous emission rate (Purcell effect)}

The motivation behind investigating slow light waveguides in the context of
enhanced spontaneous emission is that, quite generally, the relevant
contribution to the LDOS from a lossless waveguide mode is inversely
proportional to the group velocity \cite{sheng}, and so slow light modes may
lead to significant PFs. In the field of planar photonic crystal waveguides,
GF calculations~\cite{MangaPRB07} and recent measurements~\cite{HansenPRL08}
have obtained PFs greater than 30 for group velocities that are about 40
times slower than $c$. This enhancement leads to an increase in the degree
of light-matter interaction, and is important for fundamental processes such
as nonlinear optics, and for applications such as single photon sources. The
major difference with lossless photonic crystal waveguides and NIM
waveguides is that the NIM losses will likely mean that they are unlikely to
find practical application as efficient photon sources, since the photon
emission is probably dominated by non-radiative decay. Nevertheless, it
is fundamentally interesting to calculate the emission enhancements rates,
and to connect these to a measurement that would allow direct access to this
enhanced light-matter interaction regime.

\begin{figure}[t]
\includegraphics[width=0.45\textwidth]{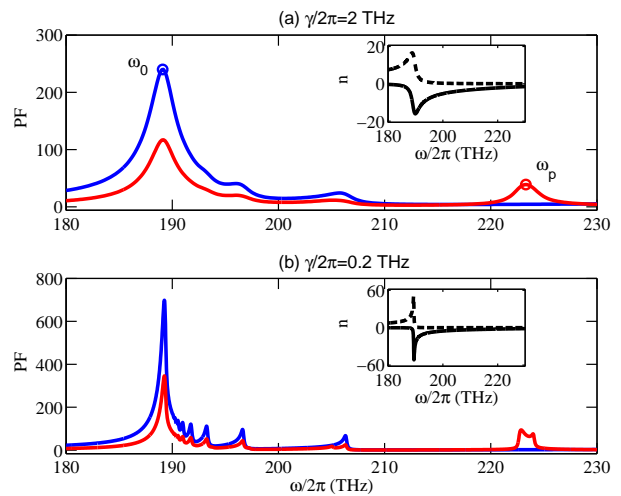}
\caption{Purcell factor as a function of frequency for $\protect\gamma /2%
\protect\pi =2\,\mathrm{THz}$ (a) and $\protect\gamma /2\protect\pi =0.2\,%
\mathrm{THz}$ (b). The dot is located at $z_{0}=h/10=28\,$nm ($0.02\,\protect%
\lambda _{0}$) above the NIM slab. The blue curves correspond to a $z-$%
polarized dipole, and the red curves correspond to an $x/y-$polarized
dipole. The inset is the refractive index, $n$, vs frequency, where the
dashed line corresponds to the imaginary part and the solid line to the real
part (which is negative throughout the entire frequency range shown).}
\label{fig:2}
\end{figure}

For our PF calculations, we first investigate the behavior of the
spontaneous emission as a function of frequency. The GF is obtained directly
using the multilayer GF technique described above. Figure~\ref{fig:2} shows
the spectral distribution of the PF, when the emitter is placed above the
slab ($z_d=z_{0} \ {\rm (see \ Fig.~1)} \ =h/10=28\,$nm) and the loss rate $\gamma /2\pi $ of the
metamaterial is 2\thinspace THz and 0.2\thinspace THz for (a) and (b),
respectively.
A height of 28\,nm
corresponds to a normalized distance of $z_{0}/\lambda _{0}\approx 0.02$,
where we can reasonably expect the dipole approximation to hold (the spatial
dependence will be shown below).

Our  material loss numbers ($\gamma /\omega
_{0}=0.01-0.001$) are close to the state-of-the-art for metamaterials, but
they are significantly greater than those used in some previous waveguide
studies, where enhanced PFs were demonstrated with $\gamma /\omega
_{0}\approx 10^{-10}-10^{-8}$~\cite{LiPEB2009}, or no loss at all~\cite%
{XuPRA2009}.
For metamaterial applications, a useful
figure of merit (FOM), is $\mathrm{FOM=-Re(n)/Im(n)}$, with a larger FOM
indicating a less lossy metamaterial. The current FOM for typical
metamaterial is of order 100 at GHz frequencies and drops to 0.5 at optical
frequencies (380 THz)~\cite{ol_32_53}. 
However, there are methods to improve these FOMs as they are not fundamental
material properties; for example, Soukoulis \emph{et al.} have 
suggested that the FOM can be improved by a factor of 5 at optical
frequencies, and 
after optimizing their \emph{fishnet design}, they have demonstrated
that the FOM can be around 10 at 380\thinspace THz (cf.~Fig.~5(c) in Ref.~%
\cite{oe_16_11147}). Recently, they also demonstrated a new design where the
FOM is about 60 at 40 THz~\cite{prl_102_053901}.
For the proposed structure with $\gamma/2\pi = 2\,$THz, we have a FOM of 0.72
at the resonance frequency, and a maximum  FOM of 26.25 at 220\,THz,
which is similar to  the state-of-the
art FOMs at optical frequencies.

The PF (Eq.~\ref{PF2}) is enhanced at the frequency of $\omega /2\pi \approx
189\,\mathrm{THz}\,(\approx \omega _{0}/2\pi )$, and the enhancement for a $z-$%
polarized dipole is larger than that for an $x$ or $y-$polarized dipole;
part of the reason that the PF is larger for the $z-$polarized dipole is
that it couples to the TM modes, which are more strongly influenced by
material losses (through $\mu $); we have verified that the TM and TE PFs
approach one another as $\gamma \rightarrow 0$, and  this trend can partly be seen
by comparing the TE and TM PFs in Fig.~4a and Fig.~4b. In the presence of the larger
loss ($\gamma /2\pi =2$\thinspace THz) the peak $F_{z}$ (TM PF) is about 240
and the peak $F_{x/y}$ is about 120. When $\gamma /2\pi $ is decreased to
0.2\thinspace THz, the PF increases significantly to $\Gamma _{z}=720$ and $%
\Gamma _{x/y}=350$. The physical origin of these large PF enhancements comes
from the slow energy velocities of the propagation modes. This can be seen
from the dispersion curves (Fig.~2 and Fig.~3), where upon close inspection,
we realize that we are obtaining the odd mode resonances, as is expected
from dipoles near the surface where the local field is larger
near the NIM surface; for example, the resonance around
207\thinspace THz corresponds to the $v_{g}\rightarrow 0$ region of the TE3
mode (cf.~ Fig.~2(c) on the complex$-\omega $ band structure).
In the complex$-\beta$ dispersion curves (Fig. 3(b)), this
same resonance is seen as the point where the two branches of the TE3 mode
split apart near 207 THz ($\beta \simeq 7\,\mu $m).
The series of peaks below 200\,THz and above
$\omega_0$ are due to the slow group velocity region of the various
odd modes, which approach one another at $\omega_0$.

In addition, we note that for the TE modes, there is a PF peak due to the
plasmon resonance around 223\,THz
that has also been highlighted elsewhere, e.g. Refs.~\cite%
{RuppinMartin2004,XuPRA2009,LiPEB2009}. What is particularly interesting, is
that for reduced losses, this resonance splits into two (cf.~Fig. 4(b));
moreover, by inspection of the band structure (cf. Fig.~3(b)), the
lower-lying peak of this pair is actually due to the bound
propagation modes (e.g., TE5), which is only visible in the complex$-\beta $ band
structure. For larger losses, these individual peaks cannot be resolved, and
one must then assume that the broadened peak near 223\thinspace THz
for $\gamma /2\pi =2$\thinspace THz, is due to a
combination of the TE SPP and bound modes contained within the light lines.
We emphasize that this resonance, which is below the SPP frequency,
is \emph{not} observable with the complex$%
-\omega $ band structure, as discussed earlier; it is alo
unique to the NIM structure.


\begin{figure}[t]
\includegraphics[width=0.4\textwidth,height=0.3\textwidth]{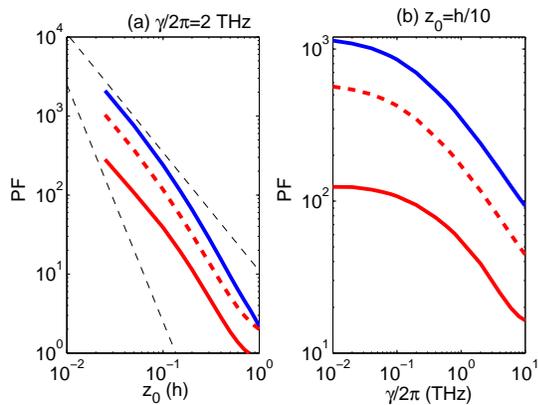}
\caption{(a) The dependence of the maximum values in the PF on position $%
z_{0}$ for a loss factor of $\protect\gamma /2\protect\pi =2\,\mathrm{THz}$%
.. \ The solid blue and red curves correspond to the PF at $\protect\omega %
_{0}$ for TM and TE modes respectively and the dashed red curve corresponds
to the TE SPP mode. The upper and lower dashed grey curves show the scaling
of $1/z_{0}^{1.5}$ and $1/z_{0}^{3}$, respectively.
Even at the smallest distances from the NIM surface ($\sim 7\,$nm),
the electrostatic limit (where a scaling of $1/z_{0}^{3}$ would occur) has still not been reached at the chosen frequencies
(see text).
(b) The maximum PFs as a
function of damping factor for the dot located at $z_{0}=h/20$. \ The curve
labelling is the same as in (a). All curves are obtained at a fixed
frequency. }
\label{fig:3}
\end{figure}

Next, the PF for different dot positions $z_{0}$ is investigated. Similar to
what happens near metal surfaces, as the dot is brought closer to the NIM
slab surface, the PF will increase rapidly and formally diverge at the
surface of the NIM slab. 
In fact, it is a straightforward exercise to show analytically that the
electrostatic $\mathbf{G}$ at the surface of a half-space lossy structure
has an imaginary part that diverges. One finds for the non-retarded terms
(quasi-static approximation)~\cite{MartinAO2001}: $\mathbf{G}(\mathbf{r}_{s},%
\mathbf{r}_{s})=\mathbf{G}^{0}(\mathbf{r}_{s},\mathbf{r}_{s})\mp \mathbf{G}%
^{0}(\mathbf{r}_{s},-\mathbf{r}_{s})(\varepsilon _{nim}-1)/(\varepsilon
_{nim}+1)$, where the minus and plus sign refer to TE and TM polarization
respectively, $\mathbf{G}^{0}$ is for free space, and $\varepsilon _{nim}$
is the permittivity of the NIM medium. \ Because ${\rm Re}\left\{ \mathbf{G}%
^{0}(\mathbf{r}_{s},\mathbf{r}_{s})\right\} $ diverges, any amount of loss
in $\varepsilon _{nim}$, no matter how small, will lead to a divergent LDOS
at the surface. Consequently, the quasi-static approximation will not work
at the surface, and in general we should consider distances significantly
larger than the emitter size if we are to employ the dipole approximation.

Figure \ref{fig:3}(b) shows the dependence of the values of the PF at
three different resconance peaks on the position $z_{0}$. Because of the expected LDOS
divergence at the surface, %
we only show the
behavior down to distances of $h/40$ ($%
z_{0}/\lambda _{0}=0.005$); we expect the dipole approximation to work to
distances of around $h/10$ ($z_{0}/\lambda _{0}=0.02$). In obtaining these
graphs we have fixed the frequency.
The PF enhancements are found to decrease as a function of distance, as
expected, but large values can still be obtained at distances ($\sim 0.4h$)
or more. \ For example, the TM peak has a PF of 10 at a distance of
100\thinspace nm from the surface. For metal surfaces, the TM SPP mode
PF scales as $1/z_{0}^{3}$ for small distances (e.g. see Ref.~\cite%
{JoulainPRB2003}), which we have also verified for our structure (see Fig.
5(a)). This behavior is due to the electrostatic scaling of the evanescent
contribution from the SPP, and for our structure, this scaling dominates for
distances of around $z_{0}\leq 0.04\,h$. However, the scaling of the
metamaterial modes is quite different: we obtain a scaling of around $1/z_{0}^{1.5}$
for the PF peak at $\omega _{0}$ and also for the TE SPP mode peak. In the limit
of $\mu =1$, we again recover the $1/z_{0}^{3}$ scaling for these modes.
Also in the case of the metamaterial, significant PFs can still be achieved
for much larger distances away from the surface, even for $z_{0}=h$.
The reason for this unexpected scaling is that the chosen resonance frequencies
have not yet recovered the electrostatic limit, even for
dipole distances as small as 7\,nm from the surface; if we choose
frequencies away from the waveguide peaks, then we indeed
get the the $1/z_{0}^{3}$ scaling from the NIM, as expected.

The dependence of spontaneous emission on the damping factor is plotted in
Fig.~\ref{fig:3}(b), which shows that increasing the damping factor
decreases the peak spontaneous emission rate, while increasing the full
width at half maximum (FWHM) of the PF resonance. However, even in the
presence of very large losses (e.g., $\gamma /2\pi =$10\thinspace THz), we
see that large PFs are still achievable. As expected, clearly there is also
a large improvement in the PF enhancement if the nominal losses can be
improved by an order of magnitude.

\subsection{Lamb shift and far-field spectrum of spontaneous emission}

The Lamb shift is another fundamental quantum effect whereby the vacuum
interaction with a photon emitter can cause a frequency shift of the emitter~%
\cite{lamb}. Cavity QED level shifts of atoms near a metallic surface have
been shown to be significant as one approaches the surface~\cite{SipePRA1985}%
. For optimal coupling, usually these are studied at high frequencies (e.g.,
$\hbar \omega >4\,$eV), so as to couple to the TM SPP resonance; as a
function of frequency, the level shift changes sign as we cross the
resonance~\cite{SipePRA1985}. For lower frequencxies, a $1/z_{0}^{3}$ van
der Waals scaling again occurs~\cite{Hinds1991}. Given the complicated modal
structures of NIM waveguides, it is not clear what the Lamb shifts will look
like, and to the best of our knowledge,  Lamb shifts have never been
studied in the context of metamaterial waveguides.

\begin{figure}[t!]
\includegraphics[width=0.44\textwidth]{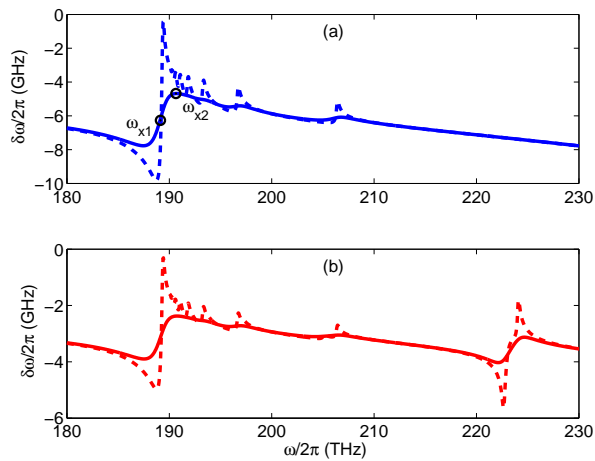}
\caption{Frequency shift due to the inhomogeneous scattering when $\protect%
\gamma/2\protect\pi = 2\, \mathrm{THz}$ (solid curves) and $\protect\gamma/2%
\protect\pi = 0.2\, \mathrm{\ THz}$ (dashed curves). The dipole strength is $%
d=50\,D$ (see text). (a) The dot polarization is perpendicular to the slab
surface (TM). (b) The dot polarization is parallel to the slab surface (TE).
}
\label{fig:lamb_shift}
\end{figure}

The QED frequency shift of the emitter can be directly calculated via the
real part of the scattered GF (Eq.~(19)). Experimental dipole moments for
semiconductor quantum dots vary from around $30\,\mathrm{D}$ (D=Debye) to
$100\,$D~\cite{silverman,stievater}, so here we adopt a realistic
dipole moment of $d=50\,$D. The results for $z-$polarization and $%
x/y-$polarization are shown in Fig.~\ref{fig:lamb_shift}(a) and Fig.~\ref%
{fig:lamb_shift}(b), respectively. Our calculations indicate that the
frequency for a dipole above the NIM slab will be significantly red-shifted
relative to vacuum, with rich frequency oscillations as one sweeps through
the NIM waveguide resonances. The Lamb shift at the $\omega_0$ resonance frequency
for different loss factors $\gamma $ are identical, and are not zero;
the nonzero shift at the main resonant frequency $\omega _{0}$ is due to the
asymmetry of the PF, namely the series of waveguide resonances
at the higher frequency end of $\omega_0$. The frequency shift at $%
\omega _{0}$ for TM modes is 6.3\thinspace GHz, and for TE modes is
3.5\thinspace GHz, and the difference between them mainly comes from TM SPP
modes at $\omega _{pe}/\sqrt{2}\approx 490\,$THz. When the loss is reduced,
the various modal contributions become more pronounced. It is
worth highlighting that this frequency shift, which is completely
unoptimized, is already comparable to some of the largest shifts reported
for the real-index photonic crystal environment, e.g., $|\delta \omega
|/\omega =4\times 10^{-5}$\cite{wang}. In normalized units, we obtain
frequency shifts around $|\delta \omega |/\omega =5\times 10^{-5}$, over a
wide frequency range. 
This ratio is even larger for
smaller distances (and larger dipole moments), however
one must watch that the dipole approximation does not breakdown.

We also remark that these NIM Lamb shift features are substantially different
to those predicted in typical metals. For example, we have calculated the
Lamb shift from a half space of Aluminium and find that the Lamb shift in
the same optical frequency regime is only -0.1\thinspace GHz, for an
identical dipole and position. \ Although closer to the SPP resonance (which is at
2780\thinspace THz), much larger values ($PF=322$ at $z_0=7\,$nm)
can be achieved,  the frequency dependence is relatively featureless, in
contrast to that shown in Fig. 6 and the spontaneous emission is very
strongly quenched in metals near the SSP resonance.

\begin{figure}[t!]
\includegraphics[width=0.44\textwidth]{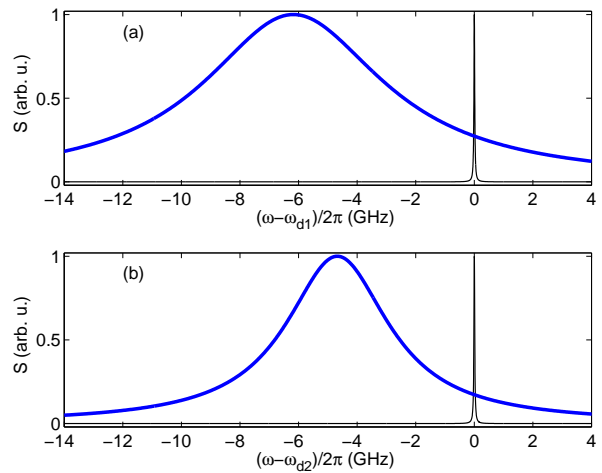}
\caption{ Detected spontaneous emission spectrum from a dot above the
metamaterial (thick blue curve), compared with the emission from the same
dot in free space (thin black curve). The dot location is $z_0=28\,$nm above
the surface and the detector position is $2\protect\lambda_0 \approx 2800\,$%
nm above the quantum dot. The dot polarization is perpendicular to the slab
surface. (a) The excition frequency is is $\protect\omega_{d1}=189.1\,$THz.
(b) The excition frequency is $\protect\omega_{d2}=190.7\,$THz.}
\label{fig:fig_4}
\end{figure}

Finally, we turn our attention to the spontaneous emission radiation
that can actually be measured. In the following, we use Eq.~(21) to investigate the emitted spontaneous emission spectrum for
two different exciton frequencies, $\omega _{d}=\omega _{d1}$ and $\omega
_{d2}$ which are indicated in Fig.~\ref{fig:lamb_shift}(a). The polarization
of the quantum dot exciton is assumed to be along $z$, and the loss factor is $\gamma
/2\pi =2\,$THz. The frequency dependence of the emitted radiation is shown in
Fig.\thinspace \ref{fig:fig_4} for the two different dot frequencies. For $%
\omega _{d}=\omega _{d1}$, the shift at peak emissions is $-6.3\,$%
GHz and the enhancement in spontaneous emission rate is 240. For $\omega
_{d}=\omega _{d2}$, the shift is $-4.7\,$GHz and the enhancement in the
spontaneous emission rate is 135. Usually, with a large dipole moment of $%
d=50$\thinspace D and a spontaneous emission enhancement on the order of $100
$, the the photon-dot interaction will enter the strong coupling regime and
the emitted spectrum will show a typical spectral doublet. \ However, since
the emitted field away from the NIM is predominantly carried away by
radiation modes, there the far field contains signatures
of the QD coupling, even to near fields; however, because
there is quenching in the far-field emission,  strong coupling is not observable in
the far field from the NIM waveguide-such effects 
could possibly be observed in
the near field given suitable detection capabilities, e.g., from a scanning
near field optical microscope. These features depend upon the properties of
the GF propagator appearing the spontaneous emission formula (Eq.~(21)).
Comparing with the free space emission spectrum, the integrated
emission in Fig.~7(a) and Fig.~7(b) is $1.8\times 10^{-4}$
and $1.6\times 10^{-3}$, respectively. Thus, the predicted
far field spectra, which obtain Purcell factor and Lamb shift signatures,
 should certainly be observable experimentally.

\section{Conclusion}

\label{Conc} In summary, we have employed a rigorous medium-dependent theory
and a stratified Green function technique, to investigate the enhanced emission
characteristics of a single photon emitter
near the surface of the NIM slab waveguide in the optical frequency regime.
The origin of the predicted Purcell factor peaks is primarily due to slow light propagation modes
which have been analyzed by calculating the complex band structure of this
waveguide. Correspondingly, we also predict a significant frequency (Lamb)
shift of the single photon emitter, with rich features that stem
from the waveguide mode characteristics. All of our predictions are based on
a realistic metamaterial model that includes both material dispersion and
loss and scales to any region of the electromagnetic spectrum; the role of
material loss and dipole position has also been investigated in detail.
It is further shown that the rich
emission characteristics at the surface can act as a sensitive and
non-perturbative probe of below-light-line waveguide mode
characteristics.
These predicted medium-dependent QED effects are fundamentally interesting
and may find use for applications in
quantum information science.



This work was supported by the National Sciences and Engineering Research
Council of Canada and the Canadian Foundation for Innovation. We thank
O. J. F. Martin and M. Wubs for useful discussions.


\clearpage

\clearpage

%
%



\begin{thebibliography}{99}
\bibitem{RuppinMartin2004} R. Ruppin and O. J. F. Martin, J. Chem. Phys.
\textbf{121}, 11358 (2004). We note that Eq.~(1) should have $4\pi$ replaced
by $6\pi$ in this paper, and $\mathbf{G}$ as defined is $\mathbf{G}^{\mathrm{%
scatt}}$.
\end{thebibliography}
\end{document}